# A framework to evaluate whether to pool or separate behaviors in a multilayer network

Running head: Decision framework for pooling or splitting behaviors

Annemarie van der Marel[1], Sanjay Prasher[1], Chelsea Carminito[1], Claire O'Connell[1], Alexa Phillips[1], Bryan M. Kluever[2], Elizabeth A. Hobson[1]*

[1] Department of Biological Sciences, University of Cincinnati, Ohio USA

[2] United States Department of Agriculture, Wildlife Services, National Wildlife Research Center, Florida Field Station, Florida USA

*Corresponding author: elizabeth.hobson@uc.edu

**Abstract**

A multilayer network approach combines different network layers, which are connected by interlayer edges, to create a single mathematical object. These networks can contain a variety of information types and represent different aspects of a system. However, the process for selecting which information to include is not always straightforward. Using data on two agonistic behaviors in a captive population of monk parakeets (*Myiopsitta monachus*), we developed a framework for investigating how pooling or splitting behaviors at the scale of dyadic relationships (between two individuals) affects individual- and group-level social properties. We designed two reference models to test whether randomizing the number of interactions across behavior types results in similar structural patterns as the observed data. Although the behaviors were correlated, the first reference model suggests that the two behaviors convey different information about some social properties and should therefore not be pooled. However, once we controlled for data sparsity, we found that the observed measures corresponded with those from the second reference model. Hence, our initial result may have been due to the unequal frequencies of each behavior. Overall, our findings support pooling the two behaviors. Awareness of how selected measurements can be affected by data properties is warranted, but nonetheless our framework disentangles these efforts and as a result can be used for myriad types of behaviors and questions. This framework will help researchers make informed and data-driven decisions about which behaviors to pool or separate, prior to using the data in subsequent multilayer network analyses.

**Introduction**

Traditional social network analysis has provided significant insight into the form and function of social systems, but sociality is often multifaceted. Including multiple types of social interactions provides a richer description of social structure (Whitehead and Dufault 1999) and can allow for better integration of multiple factors, such as spatial, temporal, and genetic relatedness along with social interactions to better explain patterns of sociality. This *multilayer network* perspective has gained recent attention because it provides a framework for combining social analyses and allows researchers to analyze sociality as one mathematical object (Barrett et al. 2012; De Domenico et al. 2013; Bianconi 2018; Silk et al. 2018; Finn et al. 2019; Beisner et al. 2020; Pereira et al. 2020). Analyzing multiple layers together can provide more comprehensive insight into the factors affecting sociality, the hidden mechanisms of a system, and social structure patterns in animal societies than analyzing any one behavioral or network type in isolation. When using multilayer network approaches, researchers must carefully consider how the network layers are assembled. Social network layers are built from associations or interactions among dyads (pairs of individuals). Determining what the layers should represent, and how to construct them, is a critical step in the formation of any single or multilayer network. In some cases, this determination is more obvious, especially when the two network layers are very different from one another (for example, genetic relatedness and social associations, Evans et al. 2020). In other cases, decisions about what behaviors to include, exclude, or treat as equivalent can be much less straightforward.

Pooling behaviors together can provide many benefits. Pooling observations of different behaviors into a single network layer can reduce data sparsity problems for some interaction types, resulting in more comprehensive networks. These pooled layers can result in better models of the real social structure which allow for better quantification of sociality. Pooling can also be used to simplify multilayer network analyses by focusing on fewer network types and reducing the number of layers used in the analysis, reducing nonindependence problems and decreasing the risk of committing Type 1 errors (Silk et al. 2013).

Although pooling behavioral data can provide benefits in multilayer analyses, it can also come with potential costs. Different behaviors may each convey important information when considered separately, and these differences may be lost if behaviors are pooled (Beisner et al. 2015; Beisner et al. 2020). The combination of two non-equivalent behaviors into a single network layer could introduce unnecessary noise into a multilayer network analysis and reduce the ability of those analyses to reach clear conclusions. Combining non-equivalent behaviors that differ in how commonly or rarely they are observed could also strongly bias the resulting network layer towards the most common behavior (Silk et al. 2013). These costs of pooling behaviors at the dyadic interaction level are especially important to consider in *multilevel analyses* where the focus is on detecting structure at different levels of social organization. Pooling seemingly-similar dyadic interactions may differentially impact more macro-scale social properties, even in cases where behaviors appear similar at the dyadic level.

Current methods for deciding whether to pool or split behaviors within a behavioral context largely fall into three main approaches found across different animal taxa: (1) unspecified decisions made at either the data collection or analysis level, (2) researcher familiarity with the biology of the study system, and (3) the strength of correlation between the behavior types at the dyadic level.

Decisions about pooling data are sometimes not well described. Details about the decision-making process of what behaviors are included in analyses, or how they may have been pooled or kept separate, are sometimes not explicitly reported in studies (*e.g.* Herberholz et al. 2003; Viblanc et al. 2016). These choices may not be reported because weighing decisions about whether to pool or split behaviors occurs at different points during a study. Decisions about network layers can be made at the time of data collection when observation protocols determine how data are coded. In these cases, it is typical for authors to report which behaviors they collected; it is less common for authors to provide a detailed description of all the behaviors that they could have collected, how those could have been subdivided into more specific categories, and why particular behaviors were categorized in certain ways. For example, in a study to identify the patterns of social ties within cichlid cooperative networks, the authors created affiliative and aggression networks and listed specific behaviors that qualified as either aggressive or affiliative; however, they did not further explain their reasoning for combining the behaviors (Schürch et al. 2010). These decisions at the data collection stage can have downstream effects on later analyses, which may be constrained by the ways data were collected. To ensure flexibility in future analytical approaches, researchers often collect a suite of behavioral interaction and association data in several contexts, such as direct affiliative or aggressive interactions, and more passive tolerance, proximity, or group associations. It is important to note that while recording more detailed observations during data collection can allow for different ways of slicing, combining, or subsetting data for future analyses, detailed data like this can also be more difficult to collect reliably, especially in cases where there are only slight differences between two desired behavioral types. In cases where many types of behavioral data are collected and coded uniquely, decisions about which types to use to construct a specific network layer come at the analysis stage. As suggested by Ferreira et al. (2020), it is important to give a detailed description of the study design as it may provide guidelines for further analyses in the same study or in other studies.

Decisions about pooling data can also be based on biological knowledge of the system. Researchers often rely on familiarity with the biology of the system to decide which behaviors "qualify" as sufficiently different to be coded separately or are similar enough to be included in the same network layer. This approach is especially common when researchers perceive two behaviors as qualitatively different types of interactions that both fall within the same social context. For example, some studies differentiate between low-level aggression and high-level aggression based on assumptions about the energetic costs or potential for injury (Oczak et al. 2014; Pierard et al. 2019; Wey et al. 2019; Beisner et al. 2020). Although the two behaviors may be coded as separate interaction types, they both fall within an agonistic social context. Researchers can also build on previous work with the same or closely related species to use knowledge of the system to make decisions about which behaviors to include or how to pool them (Munroe and Koprowski 2014; Beisner et al. 2020; Pereira et al. 2020). If the animals themselves perceive two types of behavioral interactions as socially-equivalent, biologically it would make sense to pool these two behaviors, and knowledge of the study system can be used as a rationale for making these decisions. A danger to this approach is that the study system may not be well enough understood to make these decisions in ways that align with the biological reality of how the animals themselves perceive the behaviors. In this case, pilot studies can be performed to obtain a priori knowledge about the study system, which can be helpful in the study design and analysis and may reduce the chances of making type 1 errors (Ferreira et al. 2020).

Finally, decisions about pooling data can be made using a data-driven approach. Here, researchers may use initial data analyses to evaluate whether the frequency of behaviors between individuals are correlated, whether behaviors can be condensed down to fewer types using dimension-reduction methods, or through comparing behaviors to find dissimilar or unique information. For example, in a study on the effects of perturbations in a social group on hierarchy structure in house sparrows, the authors pooled the interaction types that were correlated per behavioral context (Kubitza et al. 2015). Network layers may also be standardized by consensus ranking to identify significant vertices in separate layers (Braun 2019).

In this paper, we expand on these data-driven methods to help decide whether to pool or split data. We developed a framework to examine the implications of splitting or pooling behaviors at the dyadic level before deciding on how to construct the layers of networks in a multilayer network analysis. This framework can also be applied to simpler single-layer network analysis. We propose a three-step process for investigating the general implications of pooling versus splitting behaviors: (1) perform exploratory analyses and use prior knowledge to determine whether the behaviors belong to the same behavioral context, (2) test whether behaviors can be considered "interchangeable", and (3) test how data sparsity may affect the extent to which behaviors are interchangeable (see Figure 1). Our approach highlights how pooling or splitting behaviors may differentially affect measures of social structure across different levels of social organization (Hobson, Ferdinand, et al. 2019). We focus on how changes in relationships (formed via different types of interactions) may affect individual- and group-level social properties like strength and centrality, network properties, dominance hierarchy structure, and aggression strategies. We illustrate how this framework can be used by applying it to two types of aggressive behavior recorded in a group of monk parakeets (*Myiopsitta monachus*). Our aim is to provide guidelines for other researchers to better evaluate these implications in their own study systems.

**Methods**

***Data collection***

To illustrate our evaluation methods, we used data collected from monk parakeet social interactions. Monk parakeets are small (100-150g) neotropical parrots that exhibit the potential for cognitive and social complexity (Hobson et al. 2013; Hobson et al. 2014; Hobson and DeDeo 2015).

We collected data on several types of social interactions in a long-term captive population of monk parakeets. The parakeets (n=21 individuals) were housed at the United States Department of Agriculture Wildlife Services National Wildlife Research Center, Florida Field Station, located in Gainesville, FL, USA. Observations occurred during March 2020 (the field season was cut short due to the COVID-19 pandemic). To enable individual identification, we marked each parakeet's feathers with a unique color combination using nontoxic permanent markers (Hobson et al. 2013). We released these marked birds into a large 45x45m semi-natural outdoor flight pen and then allowed the social structure time to stabilize. Observations reported here occurred after the birds had been in the flight pen and interacting for nine days.

Observers were stationed in blinds in three locations around the flight pen to conduct observations; 3-4 observers collected observations between 09:00 and 19:00 daily. We used all-occurrence sampling (Altmann 1974) and recorded dyadic interactions using the Animal Observer v1.0 app, directly inputting the data on iPads. For this analysis, we present data collected on displacements (instances where one bird aggressively approached another bird and supplanted it from its location, sometimes via physical contact) and crowds (where one bird approached another bird which moved away before the aggressor was within striking range) during a 3-day period when the dominance structure was stable in the group. We differentiated these two behaviors because they appeared to differ in the severity of aggression: displacements could result in injuries (Hobson, pers.obs.) while crowds were by definition always non-contact aggressions.

Having 3-4 observers recording observations at the same time allowed us to conduct more comprehensive all-occurrence sampling, but often resulted in different observers logging the same interaction. To remove these duplicated observations, we summarized by the number of observations per interaction type that were observed in the same minute across each of the 3-4 observers. We filtered the observations to keep those from whichever observer recorded the highest number of observations of a certain interaction type in each minute, removing all potentially-duplicated observations from other observers. We also filtered the data to only include crowds or displacements where both the aggressor and the subject were identified.

### *Decision framework to evaluate the potential effects of pooling behaviors*

To test how pooling two interaction types may affect social properties, we followed our 3-step evaluation framework (Fig.1). First, we quantified the similarity and differences between the behaviors of interest. Second, we tested whether the behaviors can be considered interchangeable. Third, we determined the extent to which these results about the potential for interchangeability may be affected by the rarity of a behavior.

Step 1: Quantifying similarities and differences between behaviors

The first step in our framework is to examine the behaviors of interest to determine how they are similar and different. To do this, we examined (1) whether the behaviors differed in how commonly they were observed, (2) whether dyads exclusively used one or multiple behavior types in their interactions, and (3) how strongly the two behaviors were correlated.

We compared the behaviors to determine if one was more common than another by counting the total number of observed behaviors that were coded as crowds and the total coded as displacements. We compared the percent of observations that were crowds to the percent that were displacements. We then looked at how dyads used each behavior type by finding the number of dyads interacting solely by crowding, solely by displacing, or using a mix of both crowding and displacing. Finally, we quantified the correlation between the crowd network and the displacement network. We constructed both networks as directed and weighted association matrices where the strength of the association was the number of times each individual interacted with other group members. Both networks were asymmetric directed networks (one individual displacing/crowding another). We used Mantel tests to find the matrix correlation strength between crowd and displacement networks.

## Step 2: Determining whether the behaviors are interchangeable

The second step in our framework is to determine whether behaviors can be considered interchangeable. To do this, we constructed a reference model to test whether observed patterns were consistent with expected patterns (if behaviors are fully interchangeable) for a suite of social measures. We use the term *reference model* for random networks where some features are constrained to match those of an observed network (Gauvin et al. 2018; Hobson et al. *in review*). Randomizing or permuting some but not all of the structure of interactions is a common tool used in social network research (Farine and Whitehead 2015), Hobson et al., in review).

We constructed a permutation-based reference model (Hobson et al. *in review*) to test whether randomly reallocating total aggressive events by behavior type changed social properties (reference model 1, Fig. SM 1). We looked for changes in both individual- and group-level social properties. For each run of the model, we summarized the number of displacement and crowd interactions for each dyad; the sum of both interactions is the total number of interactions in an agonistic context for each dyad. We then randomly re-allocated the total number of agonistic interactions back to the two interaction types for each dyad (see Supplemental Material 1). This reference model preserves the total number of individuals in the group, which individuals interacted in an agonistic context, and number of total agonistic interactions. The reference model randomizes only the number of interactions that were categorized as displacements versus crowds (n=100 runs). Across many social properties, we compared the observed values to the values expected if our two behaviors were interchangeable using the proportion of random values that are less than the observed values. We used two-tailed tests: observed values needed to be <0.025 or >0.975 of values produced by the reference model to be considered significantly different.

## Step 3: Examining the effect of data sparsity on behavioral interchangeability

The third step in our framework is to investigate whether observed differences between two behaviors could be due to one behavior occurring much more frequently than the other, rather than simply due to a lack of interchangeability. We constructed another reference model (reference model 2) to test this by combining an initial subsampling procedure (Fig. SM 2A) followed by reallocating behaviors as we did in reference model 1 (Fig. SM 2B). As crowds were the rare interaction type in our dataset, we produced random subsamples of displacements equaling the total number of crowd events and then reallocated the total number of agonistic interactions as either crowds or displacements for each run of the model (n=100 runs). This reference model preserved the total number of individuals in the group, which individuals interacted in each agonistic context (crowd versus displacements), and the number of crowds observed for each dyad. The reference model randomized which of the total observed displacements were subsampled in each run and the number of interactions that were categorized as displacements versus crowds. Across many social properties, we compared the observed values to the model values to investigate if our subsampled data showed evidence for behavioral interchangeability. We determined whether observed values significantly differ from random values in the reference models using the proportion of random values that are less than the observed values (as for reference model 1, we assessed the difference between the observed and reference model data using two-tailed tests).

## Using the decision framework to evaluate whether to pool behaviors

We use each step in our framework to evaluate the strength and consistency of evidence to make an overall decision about whether to pool behaviors together. In step 1 we assess similarity in behavioral use: we have preliminary evidence that behaviors could be pooled, if 1) there is no difference in how common the behaviors are, 2) at least some proportion of the total dyads use both behavior types to interact, and 3) the behaviors are correlated. If none of the dyads use both behavior types and the behaviors are not correlated, then there is relatively strong evidence that behaviors should not be pooled. In step 2 we assess whether behaviors are interchangeable: if observed data for each behavior produces summary measures that fall within the expected range of values generated by our reference model, then the behaviors are clearly interchangeable, and pooling is strongly justifiable. Otherwise, divergence from these distributions suggests that the observed behavior may need to be considered separately. If the two behaviors occur approximately equally, then only step 2 needs to be performed. In step 3 we assess whether behaviors can be interchangeable by controlling for the rarity of one of the behaviors: if subsampled data for the more common behavior produces similar results to the observed data for the less common behavior this is evidence of two things. First, it provides supporting evidence that behaviors can be pooled, and shows that the differences in the summary measures are likely due to data availability rather than a biological distinction between the behavior types. Second, it illustrates how the particular summary measures chosen may be affected by or susceptible to the availability of data. In cases where separated behaviors produce different summary measures in step 2, overlapping distributions in step 3 provide evidence that differences in results may be due simply to differences in sparsity and it may be reasonable to pool behaviors. However, if the sub-sampled data in step 3 produces different results, this indicates that any earlier differences cannot be attributed to data availability alone and the behaviors should not be pooled.

### *Comparison of observed values with reference model distributions*

We illustrated the use of our framework by comparing summary measures reflecting various social properties of our observed data to the range of values expected from the reference model distributions. We used several micro- and macro-scale social properties: (1) individual-level network measures (out-strength, betweenness, eigenvector centrality), (2) group-level network properties (average path length, efficiency), (3) dominance hierarchy measures (linearity, steepness, triangle transitivity), and (4) social dominance patterns. For each measure, we compared the observed value to those produced by our reference models. Evidence supporting pooling behaviors accrues when observed summary measures overlap with those expected by our two reference models; evidence against pooling accrues when observed values fall outside the distribution of values from our reference models.

### Testing effects of pooling on individual-level properties

We chose three individual-level social properties: out-strength, betweenness, and eigenvector centrality. We selected these measures because they represent biologically relevant aspects of social networks and are not by definition affected simply by network sparsity. We checked our choice of network measures by referring to the decision tree described in Sosa et al. (2020) for weighted and directed networks. *Out-strength* is calculated as the sum of the weight of outgoing edges and is a measure of the frequency of an individual's interactions (an individual's social activity). An individual with a high out-strength value is an individual responsible for many aggression events. *Betweenness centrality* measures the extent to which individuals are central for information flow in a network and is calculated as the number of times the node in question was included in all possible shortest paths between two nodes (Sosa et al. 2020). *Eigenvector centrality* is a measure of each individual's influence on the entire network, where the importance of an individual is dependent on the importance of other individuals in the network and is calculated by linearly transforming the adjacency matrix (Sosa et al. 2020). Individuals have high eigenvector centrality scores when they are connected to other well-connected individuals. We calculated all measures using the igraph package v1.2.5 (Csárdi and Nepusz 2006).

For each measure, we compared each individual's value using the crowd data to the value using the displacement data. At the group level, we quantified the correlation strength across all individuals for their crowd and displacement values. A strong correlation would indicate that individuals that have high values in the crowd network also have high values in the displacement network, while individuals that had low values in one network also had low values in the other network. We used the same approach to find the group-level correlation between crowd and displacement values for each run of the reference models.

## Testing effects of pooling on group-level properties

We chose two network summary measures to compare our observed networks with our permuted reference models: average path length and efficiency. Both measures provide insight into information transfer in a network. *Average path length* measures how interconnected a network is and is calculated as the average of the shortest path between all pairs of nodes in the network. We measured average path length using the function 'mean_distance' in the igraph package v1.2.5 (Csárdi and Nepusz 2006). *Efficiency* is a variation of *effective information* which is normalized to account for network size. *Effective information* reflects the noisiness or level of certainty of a system (Hoel et al. 2020) by subtracting the Shannon's entropy of a node's uncertainty (uncertainty of the edge out-weights) from the entropy of the network's uncertainty (distribution of uncertainty of the in-weights across the network) (Klein and Hoel 2020). A high value of effective information indicates that the relationships (e.g., interactions or associations) in a network are informative or are more certain (Klein and Hoel 2020). In our system, effective information (via its normalized form, efficiency) measures the noisiness in the patterns of aggression interactions among individuals. We used the package 'einet' v 0.1.0 (Byrum et al. 2020) to calculate efficiency. Values of efficiency closer to 1 mean that future states can be explained by the current one (future states are the same as the current one), whereas values closer to 0 mean that future states can have a probability of $1/n$ (completely noisy or degenerate) (Hoel et al. 2013).

<u>Testing effects of pooling on dominance hierarchy structure</u>

We used three measures of dominance hierarchy structure: linearity, steepness, and triangle transitivity. Multiple measures may result in a better description of the dominance hierarchy structure as they measure different aspects and allow for intra- and interspecific comparisons (Norscia and Palagi 2015). *Linearity* is a measure of the consistency of dyadic aggression in a hierarchy: in a strictly linear hierarchy (h' = 1), dominant individuals beat all lower-ranked individuals in aggressive contests (Landau 1951). We measured linearity using the 'h.index' function with 1000 randomizations in the R package EloRating v0.46.11 (Neumann and Kulik 2020). *Steepness* reflects the likelihood of winning a dominance encounter with adjacently ranked individuals. When a hierarchy is steep ($D_{ij}$ = close to 1) then more dominant individuals will always win against subordinate individuals, whereas when a hierarchy is shallow, subordinates may win from time to time as well. Steepness is measured as the slope of the regression line between rank order and the normalized David's scores using the dyadic dominance index which we quantified using the 'getStp' function in the R package steepness v0.2-2 (Leiva and de Vries 2014). *Triangle transitivity* is another measure of the orderliness of hierarchies, but one that focuses on transitive relationships among groups of three individuals. Triangle transitivity calculates the proportion of orderly triads compared to the proportion that are disorderly (e.g. cyclic triads, where A wins over B, B over C, and C wins over A) (Shizuka and Mcdonald 2012; Shizuka and McDonald 2015). Transitivity and linearity are similar when all dominance relationships are known but often some dyadic relationships are unknown (due to individuals avoiding interacting with specific individuals, sampling effort, etc.). Triangle transitivity is less sensitive to these null dyads than linearity. We measured triangle transitivity using the function 'transitivity', which calculates the scaled proportion of transitive triads (0 is the random expectation and 1 is maximum transitivity, no cycles) in the R package EloRating v0.46.11 (Neumann and Kulik 2020).

## Testing effects of pooling on group-level social dominance patterns

Finally, we assessed the overall *social dominance pattern* individuals used to direct aggression. The pattern type indicates how aggression is structured by relative rank differences between the aggressors and the subjects of aggression. Potential rank-structured social dominance patterns are the *downward heuristic* (individuals aggress against any lower-ranked individuals), *close competitors* (individuals mainly aggress against individuals that are just below them in rank), and *bullying* (individuals mainly aggress against individuals that are much lower in rank, see Hobson, Mønster, et al. 2019). We assessed social dominance patterns using the R package "domstruc" (Mønster, Hobson, & DeDeo, currently available at

https://github.com/danm0nster/domstruc).

### *Data availability and protocols*

All measures were quantified for crowds only, displacements only, pooled aggression (crowds and displacements), and for each run of the two reference models. We used R v4.0.0 (R Core Team 2020) for all our analyses and the packages 'Beanplot' v1.2 (Kampstra 2008) and 'ggplot2' (Wickham 2016) to make our figures. All data and code for running the analyses and generating the figures are available on GitHub (https://github.com/annemarievdmarel/pool-

separate-behaviors, van der Marel et al. 2020). All animal-related activities were approved by the University of Cincinnati IACUC protocol #AM02-19-11-19-01 and the National Wildlife Research Center Quality Assurance #3203.

## Results

### *Quantifying similarities and differences between behaviors*

We observed a total of 1215 agonistic interactions (160 crowds and 1055 displacements) over three days (23.5 hours of observation, 82.2 person-hours). Crowds were much rarer than displacements and accounted for only 13.2% of aggressive interactions. Of the 420 total possible directed dyads, 48.3% interacted agonistically (203 directed dyads). Within directed dyads, crowds and displacements did not occur equally: a small number of directed dyads only crowded (5.9%), a larger proportion of directed dyads both crowded and displaced (35.5%), while the majority of agonistic dyads interacted only with displacements (58.6%). For directed dyads that interacted agonistically in some way, we observed 0.78 ± 1.63 crowds per dyad (mean ± SD, range 0-13) and 5.20 ± 9.31 displacements per dyad (range 0-86); combined across crowds and displacements we observed 5.99 ± 10.49 agonistic events per dyad (range 1-93). The observed number of crowds and displacements in directed dyads were strongly correlated (mantel test: $r_s$ = 0.50, $p$ < 0.001). These results provide initial support for pooling behaviors, allowing us to move to steps 2 and 3 of our decision framework.

### *Effects of pooling on individual-level properties*

The observed crowd and displacement networks were significantly correlated for both out-strength and eigenvector centrality ($r_s$ = 0.88, $p$ < 0.001 and $r_s$ = 0.72, $p$ < 0.001, respectively; Figure 2) showing that individuals have similar social properties in both networks. The observed crowd and displacement networks were not correlated for betweenness centrality ($r_s$ = 0.08, $p$ = 0.75; Figure 2): individuals that were important for information flow in the crowd network were not important in the displacement network. These results provide mixed evidence, but mostly provide support for pooling behaviors as 2 out of 3 social properties were correlated (individuals have the same functional role in both networks for the correlated measures). When we compared how these observed correlation strengths compared to those produced by our two reference models, we found that the observed values of the correlation strength between crowds and displacements of out-strength and eigenvector centrality fell within the distribution of both reference models (out-strength: ref. model 1 $p$ = 0.42, ref. model 2 $p$ = 0.25, and eigenvector centrality: ref. model 1 $p$ = 0.27, ref. model 2 $p$ = 0.07; Table 1). These results suggest that both out-strength and eigenvector centrality are robust to random re-allocation of events into different behavioral types as well as random subsampling and re-allocation (Figure 2c). The observed value of betweenness centrality overlapped the runs of reference model 1 ($p$ = 0.32) but not the runs of reference model 2 ($p$ = 0.02) suggesting that betweenness centrality is not robust to subsampling and re-allocation of aggression events (Table 1; Figure 2c).

### *Effects of pooling on group-level properties*

When we analyzed group-level properties using average path length and efficiency, we found mixed results. Average path length for observed displacements only and for the observed pooled behaviors were both shorter than when we calculated average path length using only observed crowd data (Figure 3). When we compared the average path length for the observed pooled data to the reference models, we found that the pooled average path length was lower than both reference models ($p$ < 0.001 for all; Table 1). When we compared the unpooled observed data to the reference models, we found that the observed average path length for crowds fell within the distribution of reference model 1 ($p$ = 0.42) and reference model 2 ($p$ =

0.07), but average path length for displacement data was shorter than path lengths produced by reference model 1 ($p$ = 0.01; Table 1).

Efficiency for observed crowd data was higher than the displacement only and pooled observed data (Figure 3). When we compared efficiency for the observed pooled data to the reference models, we found that the pooled efficiency was significantly lower than both reference models ($p$ < 0.001 for both; Table 1). When we compared the unpooled observed data to the reference models, we found that observed crowd efficiency was higher than reference model 1 efficiencies ($p$ < 0.001) but that observed displacement efficiency was lower than reference model 1 efficiencies ($p$ < 0.001). Observed crowd efficiency overlapped with reference model 2 values ($p$ = 0.32; Table 1).

### *Effects of pooling on dominance hierarchy structure*

Results for our analyses of dominance hierarchy structure were largely consistent across all measures (Figure 4). Linearity and steepness values of the observed pooled aggression were higher than those produced by either crowds or displacements alone. Comparing observed crowds to observed displacements showed that the group's hierarchy was less linear and shallower for crowd data compared to displacement data. When we compared linearity and steepness for the observed pooled data to the reference models, we found that the observed data was significantly higher than the reference models ($p$ < 0.001 for all; Table 1). When we compared the unpooled observed data to the reference models, we found that neither crowds or displacements overlapped with the reference model 1 distribution: crowds had lower linearity ($p$ < 0.001) and steepness ($p$ < 0.001) while displacements had higher linearity ($p$ < 0.001) and steepness ($p$ < 0.001) values compared to reference model 1 (Table 1). Once we controlled displacements for rarity, the observed crowd values fell within the range of reference model 2 runs ($p$ = 0.1 and 0.18 for linearity and steepness, respectively; Table 1).

Observed crowds had higher triangle transitivity than either observed displacements or pooled observations (Figure 4). Triangle transitivity for pooled observations overlapped with the distributions of both reference models ($p$ ref. model 1 crowds = 0.44, $p$ ref. model 1 displace = 0.4, $p$ ref. model 2 displace = 0.29; Table 1). Similarly, both observed crowds and observed displacements fell within the distribution of transitivity values when behaviors were randomly re-allocated (reference model 1: $p$ = 0.08 for crowds and $p$ = 0.45 for displacements) and when displacements were subsampled and re-allocated (reference model 2: $p$ = 0.29; Table 1).

### *Effects of pooling on group-level social dominance patterns*

Rank-structured social dominance patterns in the observed datasets (crowds, displacements, and both behaviors pooled) were all consistently categorized as a bullying strategy (where individuals preferentially aggress against others ranked much lower than themselves). When we randomly re-allocated events as crowds or displacements (reference model 1) we found that almost all runs were also categorized as showing a bullying strategy (99% of the crowd reference model 1 runs and 100% of the displace reference model 1 runs) (Figure 5). When we subsampled displacements and then re-allocated the behaviors (reference model 2), we found a different pattern: 51% of the runs were categorized as having a bullying strategy, the remaining 49% of runs showed evidence of a basic downward heuristic (where individuals aggress indiscriminately towards any individual ranked below itself) (Figure 5).

**Discussion**

We developed a framework to examine the implications of splitting or pooling potentially-related behaviors prior to determining how to construct networks in multilayer network analyses. Our approach considers general features of the behaviors and whether the behaviors belong to the same behavioral context (step 1), whether behaviors could be considered interchangeable (step 2), and whether behaviors that are subsampled to match the frequency of the rarer behavior type could then be considered interchangeable (step 3). Reference model 1 (step 2) demonstrates how the effects of analyzing behaviors separately versus pooled can affect the summary measures and reference model 2 (step 3) shows how apparent differences in the observed dataset could be erased by randomly re-allocating behaviors to each behavior type. Our approach will help researchers better weigh their options and the potential implications of deciding how to analyze multiple behavioral interaction types, especially when they differ in how commonly they are observed. Taken together, we concluded that it is likely reasonable to pool the two behaviors into a single agonistic network for future multilayer network analyses (Table 1).

We discuss the implications of our framework below. We are unable to use these data for more than assessing a single snapshot of monk parakeet sociality due to the drastically shortened field season, so we refrain from biological interpretations of the parakeet social structure and focus on the decision about how to handle data from two similar behavior types. Future work, once we can gather more long-term data, will focus on these biological interpretations.

Initial analyses within our decision framework (Fig. 1, Step 1) provided support for the potential for pooling the two behaviors because we identified similar patterns between crowd and displacement data. We found that dyads performed a mix of the agonistic behaviors, indicating that the behaviors were unlikely to be used in different behavioral contexts. More generally, if there is zero overlap of dyads performing both types of behavior, the behaviors should most likely not be pooled into a single network. However, there is no clear cut-off for when researchers should or should not pool the two behaviors if some, but not a majority, of dyads use both behaviors (future work in our group will address this question more directly). We also found that the network of crowd interactions and the network of displacement interactions were strongly correlated, providing more evidence for pooling the two. A simple test for the correlation strength between the occurrence of behaviors at the dyadic level can provide an indication of whether behaviors should be pooled or considered separately but our framework tests for the implications more directly at the network level and can help researchers better evaluate these decisions when the correct choice is not obvious. Finally, we found that the two behaviors did not occur at equal frequencies, which we addressed using two different reference models in steps 2 and 3 of our framework.

Our examination of whether crowd and displacement behaviors could be considered interchangeable (Fig. 1, Step 2) showed mixed results for the question of whether the two behaviors should be pooled. For some measures (efficiency, linearity and steepness), the observed data for each behavior (both when considered separately and pooled) did not fall within the range of the first reference model. This differentiation can be evidence that each behavior should be considered separately. However, when we investigated further, we found that these results could be due to differences in how commonly each behavior was observed (Fig. 1, Step 3). When we subsampled our commonly-observed behavior (displacements) to match the frequency of our rarely-observed behavior (crowds) and then reallocated the behaviors, we found that summary measures of the behaviors separately generated from this reference model produced similar results to the observed crowd data (efficiency, linearity and steepness). These results provide evidence that the indications against pooling (from reference model 1) could be due simply to data skewed by rarity rather than biological differences between the two behaviors. Yet, the pooled data fell outside the reference model distributions for average path length, efficiency, linearity, and steepness, which suggests that pooling all the data together is different than reallocating and analyzing behaviors separately and shows that the full dataset seems to have more information/structure than splitting into a crowd or a displacement network allows us to detect. Three options exist for studies where datasets include one behavior that is rarer than another: 1) the rare interaction type can be excluded from analyses, 2) data can be pooled across behavior types, or 3) a different summary measure that is less susceptible to data sparsity can be used.

Our analyses also show that different types of interactions may affect individual- and group-level social properties distinctly. A rare behavior can result in different network outcomes: if we only used crowd data to summarize parakeet social structure, we would have concluded that monk parakeets have a less linear, more shallow dominance hierarchy, slower information transfer but more informative interactions, and different individuals transferring information than if we used the more common displacement data (Fig. 2-5). These results demonstrate that it is important to gain *a priori* knowledge about the study system and perform exploratory analyses to make appropriate decisions in the study design, which reduces the chance of type 1 errors (Ferreira et al. 2020) and diminishes 'metric hacking' (Webber et al. 2020). Choosing the right social measures is not only essential to diminish metric hacking but also to appropriately reflect the properties of the dataset. For example, triangle transitivity should be chosen as a hierarchy structure measure in sparse datasets as it is less susceptible to data sparsity than linearity and steepness (Klass and Cords 2011; Shizuka and Mcdonald 2012; Shizuka and McDonald 2015).

The aggression strategy comparisons show that the observed strategies were consistent with strategies in randomly re-allocated events (reference model 1) and that strategy type was robust to and preserved despite these randomizations suggesting that we could pool the behaviors.

However, the strategies were less robust to manipulations where data were subsampled before being re-allocated (reference model 2: part downward heuristic and part bullying strategy). The downward heuristic strategy inferred from reference model 2 is similar to the results obtained from an earlier study with the captive population of monk parakeets (Hobson and DeDeo 2015). The difference between the observed crowd strategy and the result from reference model 2 shows that this difference cannot be explained by data sparsity alone and shows that the complexity of the social dominance pattern degrades when subsampled in reference model 2 (i.e., we lose the signal of the more complex bullying strategy and detect only a simpler downward heuristic strategy).

Our framework for examining the implication of splitting or pooling behaviors is comparable to the reducibility analysis, which can be performed after the construction of a multilayer network to analyze whether any layers are redundant (De Domenico et al. 2015). The reducibility analysis measures the number of layers that can be aggregated without losing any structural information of the multilayer network. Both this method and our framework approach can be used to reduce the number of layers. However, our framework can also be implemented in monolayer social network analysis and other analyses where one has the decision to pool or split behaviors within a behavioral context. Thus, the two methods could complement one another: our framework provides insight into which behaviors to include within a behavioral context and should be implemented prior to further analyses, whereas the reducibility analysis expresses which layers are redundant and could be aggregated and can be used as a latter step in multilayer network analysis. For example, in two studies, the authors used the reducibility analysis to analyze whether any layer in a multilayer network analysis was redundant, however they did not specify how they chose to include behaviors within a behavioral context as a specific layer (Smith-Aguilar et al. 2019; Pereira et al. 2020).

In this study, we showed how a data-driven approach can be used to decide whether to pool or keep behaviors separate by applying it to parakeet social interactions. Researchers can use this framework to investigate the potential implications of pooling or splitting behaviors in their own datasets. Although we studied the general pattern of dyadic agonistic relationships and individual- and group-level social properties, our framework can be used for any behavior (affiliative, agonistic, etc.) and for any type of analyses where the researcher must make a choice about whether to pool or split behaviors, especially when the study question deals with describing or testing aspects of multilevel sociality. We expect these approaches to be especially useful in study systems with less-documented social processes, where relying on extensive knowledge of the study system to make decisions about which behavioral types are sufficiently similar or different may be difficult.

## Acknowledgements

This research was supported in part by the U.S. Department of Agriculture, Animal and Plant Health Inspection Service, Wildlife Services, National Wildlife Research Center. We thank the staff at the Florida Field Station, especially Eddie Bruce, John Humphrey, Eric Tillman, Danyelle Sherman, and Amber Sutton for their assistance and support. We thank the handling editor and two anonymous reviewers for their constructive feedback.

## Author contributions

E. Hobson and A. van der Marel designed the study; B. Kluever provided study support and logistical aid; E. Hobson, A. van der Marel, C. Carminito, C. O'Connell, and A. Phillips collected data, E. Hobson; A. van der Marel, and S. Prasher conducted the analyses; E. Hobson, A. van der Marel, C. Carminito, C. O'Connell, and S. Prasher wrote the paper; all authors provided comments on the drafts.

**Table 1.** Evidence for pooling or splitting crowds and displacements for each of the social metrics. Reference model 1 (step 2) demonstrates how the effects of analyzing behaviors separately versus pooled can affect the summary measures and reference model 2 (step 3) shows how apparent differences in the observed dataset could be erased by randomly re-allocating behaviors to each behavior type. We used two-tailed tests: observed p-values needed to be <0.025 or >0.975 of values produced by the reference model to be considered significantly different. Significant different values support splitting while non-significant values support pooling.

| Social property | Behavior | Reference model 1 | | Reference model 2 | | Overall |
|---|---|---|---|---|---|---|
| | | p-value | supports | p-value | supports | |
| **Out-strength** | | 0.42 | pooling | 0.25 | pooling | pooling |
| **Betweenness centrality** | | 0.32 | pooling | 0.02 | splitting | mixed |
| **Eigenvector centrality** | | 0.27 | pooling | 0.07 | pooling | pooling |
| **Average path length** | crowd | 0.42 | pooling | | | |
| | displace | 0.01 | splitting | 0.07 | pooling | pooling |
| **Efficiency** | crowd | < 0.001 | splitting | | | |
| | displace | < 0.001 | splitting | 0.32 | pooling | pooling |
| **Linearity** | crowd | < 0.001 | splitting | | | |
| | displace | < 0.001 | splitting | 0.1 | pooling | pooling |
| **Steepness** | crowd | < 0.001 | splitting | | | |
| | displace | < 0.001 | splitting | 0.18 | pooling | pooling |
| **Triangle transitivity** | crowd | 0.08 | pooling | | | |
| | displace | 0.45 | pooling | 0.29 | pooling | pooling |

**Figure captions**

**Figure 1**: A 3-step decision tree showing the process of evaluating whether to pool or split behaviors for multilayer network analysis

**Figure 2**: Scatterplots and distributions of matrix correlations of three individual-based metrics for observed values and reference models between crowds and displacements: out-strength, betweenness centrality, and eigenvector centrality. Figure a) shows the scatterplot of reference model 1 and b) of reference model 2. c) Distribution of the correlation strength between crowds and displacements in reference model 1 (dark blue) and reference model 2 (light blue) for out-strength, betweenness centrality, and eigenvector centrality. Observed values are indicated in red. The correlations were significant for out-strength and eigenvector centrality but not for betweenness centrality. The correlation strengths for the observed data fell within the reference model data (except for betweenness centrality and reference model 2). Overall, these results suggest pooling of the two behaviors.

**Figure 3**: Two network measures in observed and reference model data for agonistic interactions in monk parakeets: a) average path length and b) efficiency. Observed values are indicated in purple and the distributions show values from the reference model runs. The black solid lines that cross each bean are the average lines for each reference model distribution. The observed value falls within the distribution of the reference model runs when $p < 0.05$ (noted as *) and falls outside the range when $p > 0.05$ (noted as NS). Subsampling and re-allocating of data in reference model 2 erased the difference between observed and reference model 1 data, indicating we could pool the behaviors.

**Figure 4**: Three hierarchy measures in observed and reference model data: a) linearity, b) steepness, and c) transitivity. Observed values are indicated in red and the distributions show

values from reference model runs. The observed value falls within the distribution of the reference model runs when $p < 0.05$ (noted as *) and falls outside the range when $p > 0.05$ (noted as NS). The black solid lines that cross each bean are the average lines for the distributions. For linearity and steepness, results from reference model 1 suggest we should split the 2 behaviors as the observed data falls outside of the reference model distribution. But data from reference model 2, suggest that those results are due to data sparsity. Transitivity is not susceptible to data sparsity. We could pool the two behaviors.

**Figure 5**: Aggression strategies in observed data and reference model datasets. The observed strategy for 'crowd only', 'displace only', and both behaviors pooled was the bullying strategy. Of the 100 runs, 99% of the crowd reference model 1 runs, 100% of the displace reference model 1 runs, and 51% of the subsampled reference model 2 were consistent with the observed pattern. Results from reference model 1 suggest we could pool the behaviors, whereas the difference between the observed data and the result from reference model 2 shows that this difference cannot be explained by data sparsity alone.

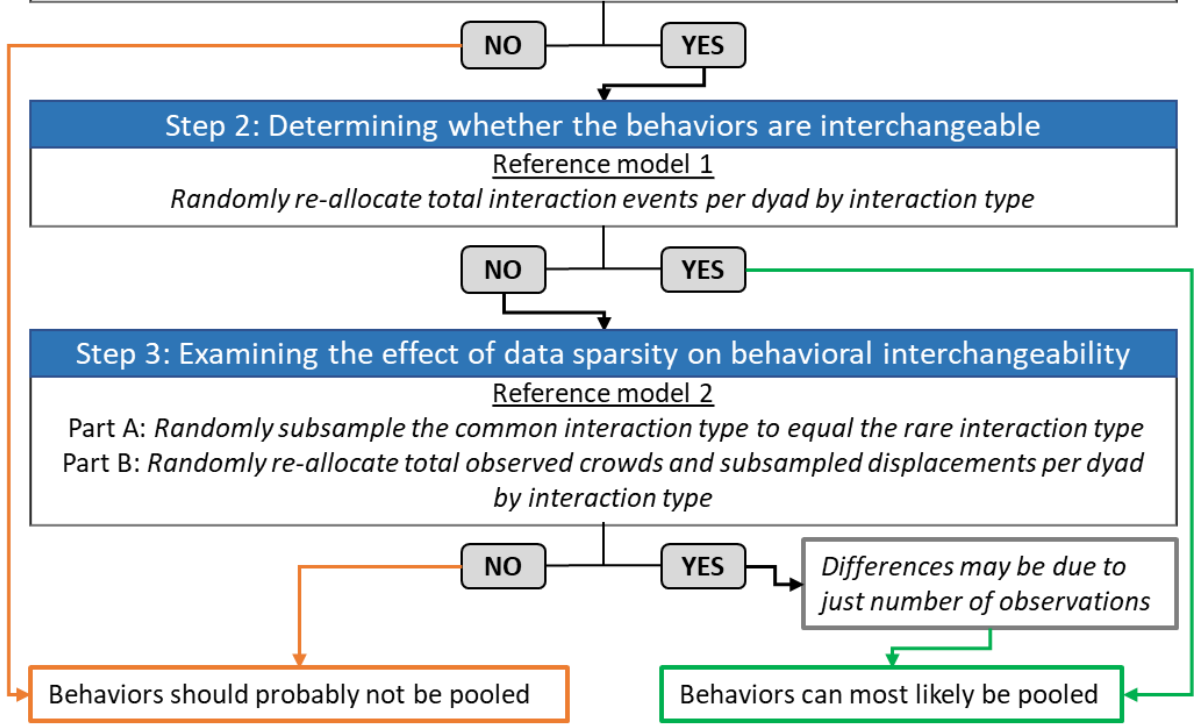

**Step 1: Quantifying similarities and differences between behaviors**

- How common are the behaviors?
*Find percent of observations that were crowds or displacements*
- Do dyads interact using a mix of the behavior types?
*Find percent of dyads which interact with only a single behavior type*
- Are the behaviors correlated?
*Find the matrix correlation strength between crowd and displacement networks*

NO    YES

**Step 2: Determining whether the behaviors are interchangeable**

Reference model 1
*Randomly re-allocate total interaction events per dyad by interaction type*

NO    YES

**Step 3: Examining the effect of data sparsity on behavioral interchangeability**

Reference model 2
Part A: *Randomly subsample the common interaction type to equal the rare interaction type*
Part B: *Randomly re-allocate total observed crowds and subsampled displacements per dyad by interaction type*

NO    YES    *Differences may be due to just number of observations*

Behaviors should probably not be pooled

Behaviors can most likely be pooled

|  | **out strength** | **betweenness** | **eigenvector** |
|---|---|---|---|

**a) Scatterplot reference model 1**

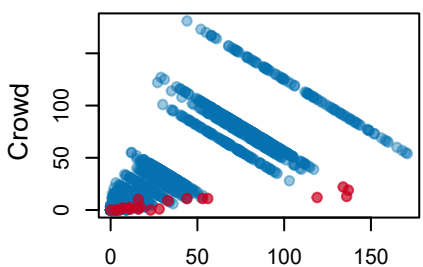 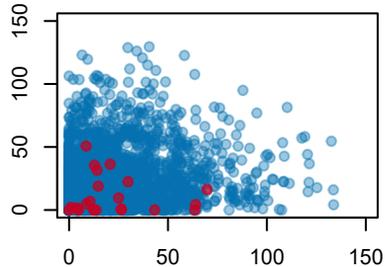 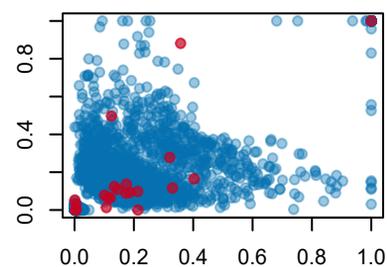

**b) Scatterplot reference model 2**

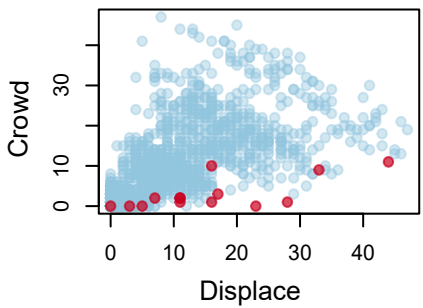 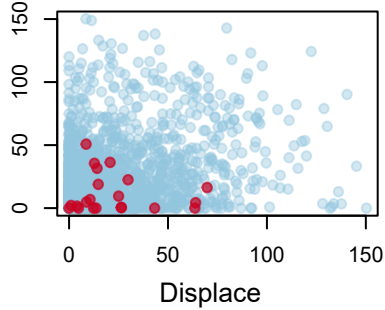 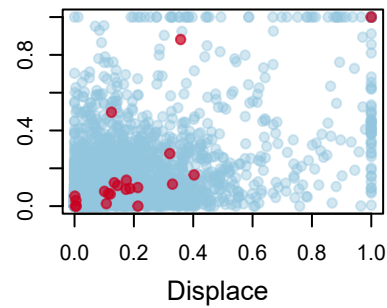

Displace

**c) Distribution of matrix correlation strength**

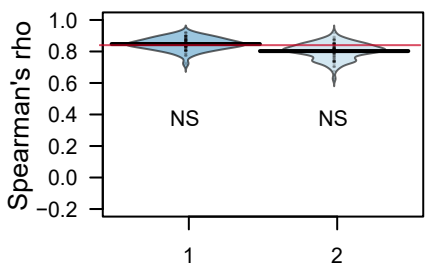 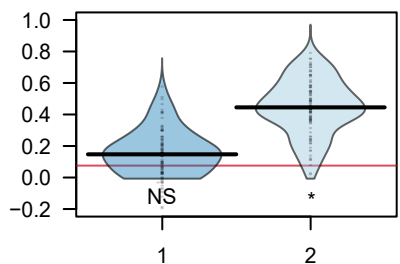 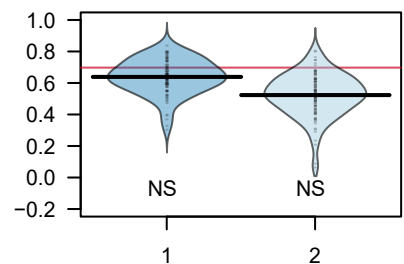

● observed values — reference model 1 — reference model 2

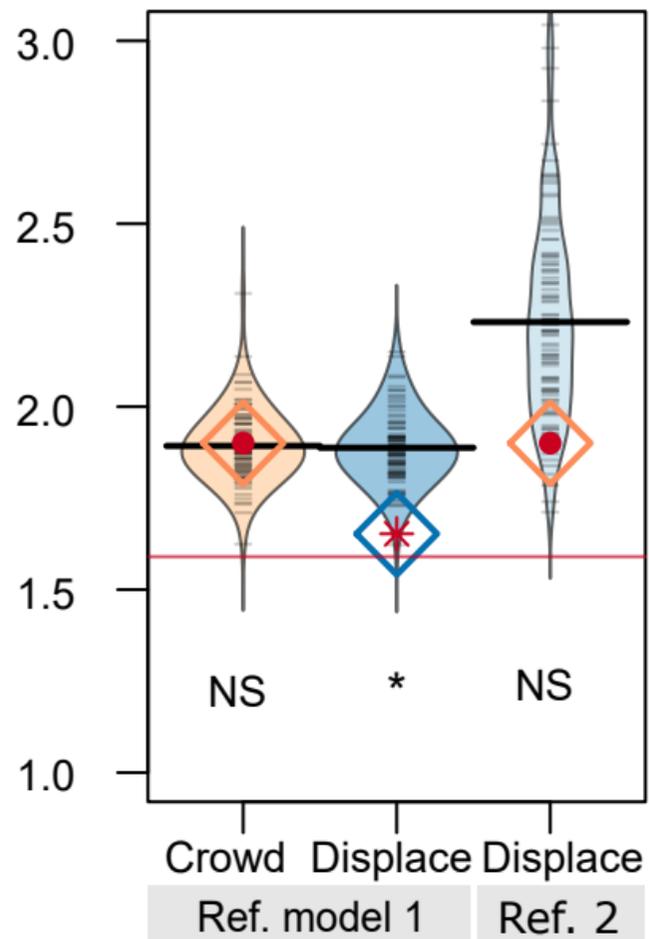

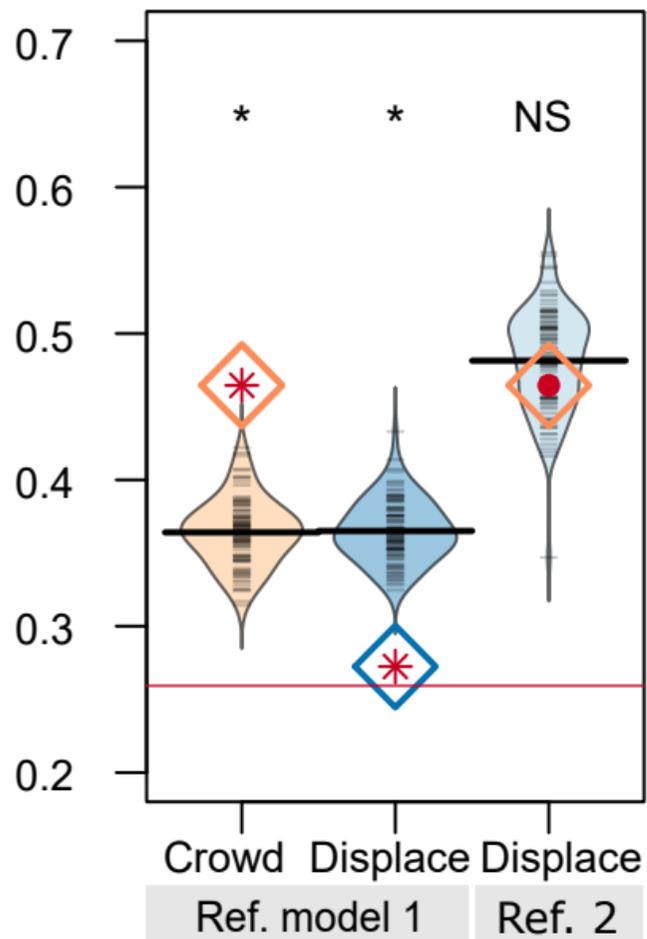

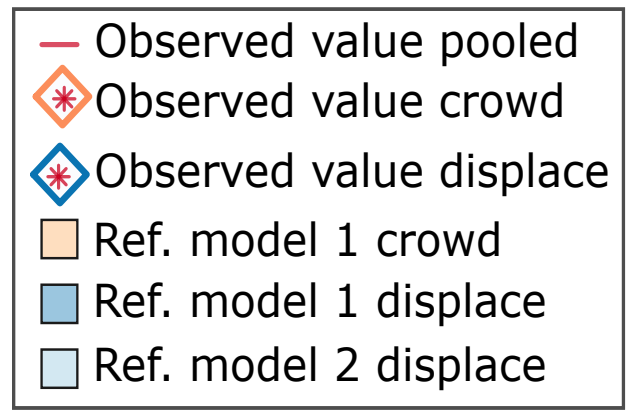

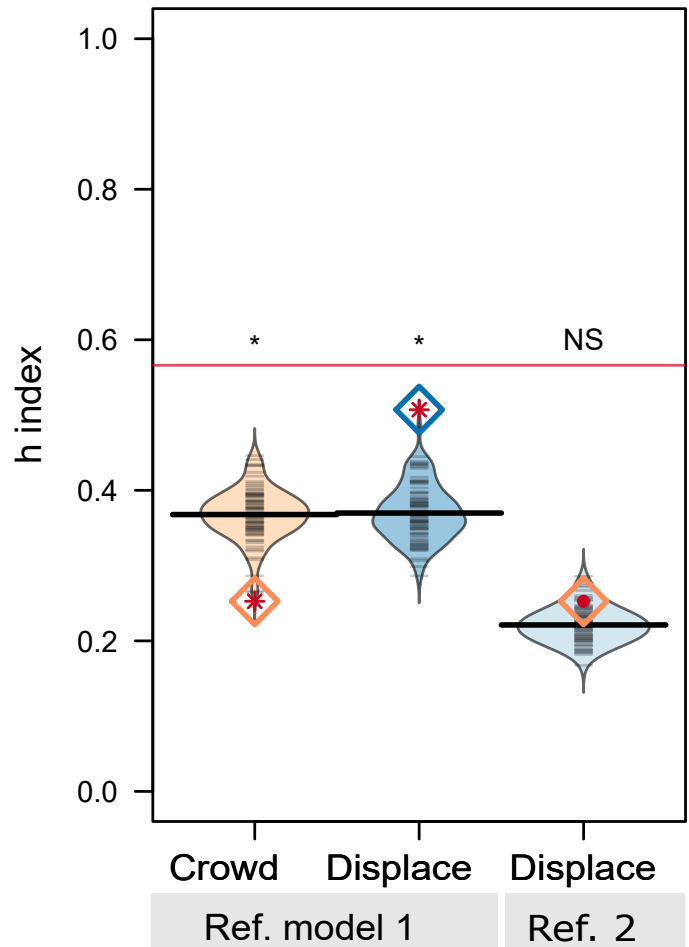

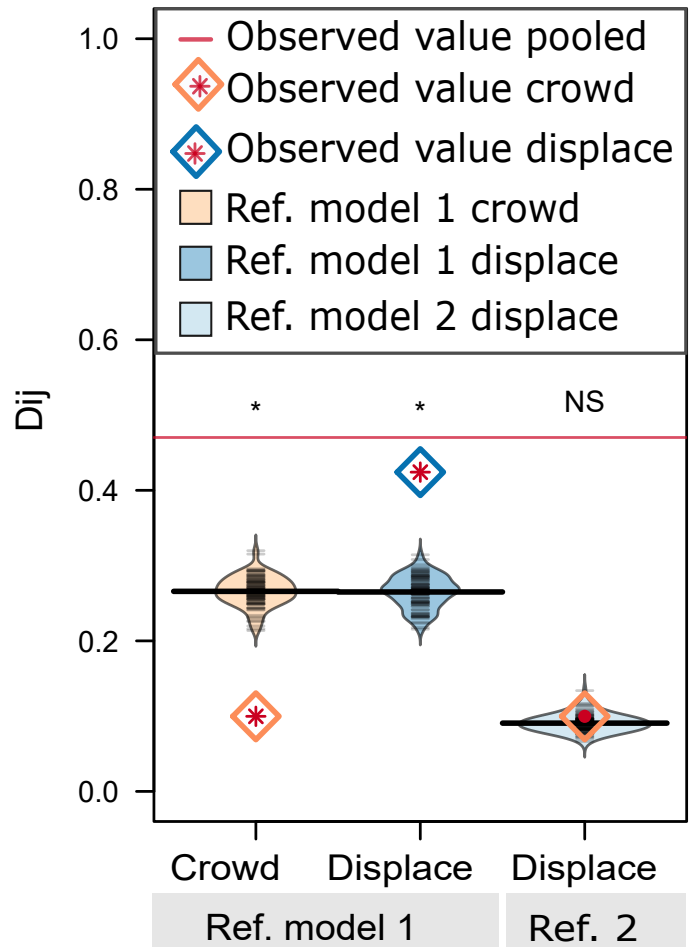

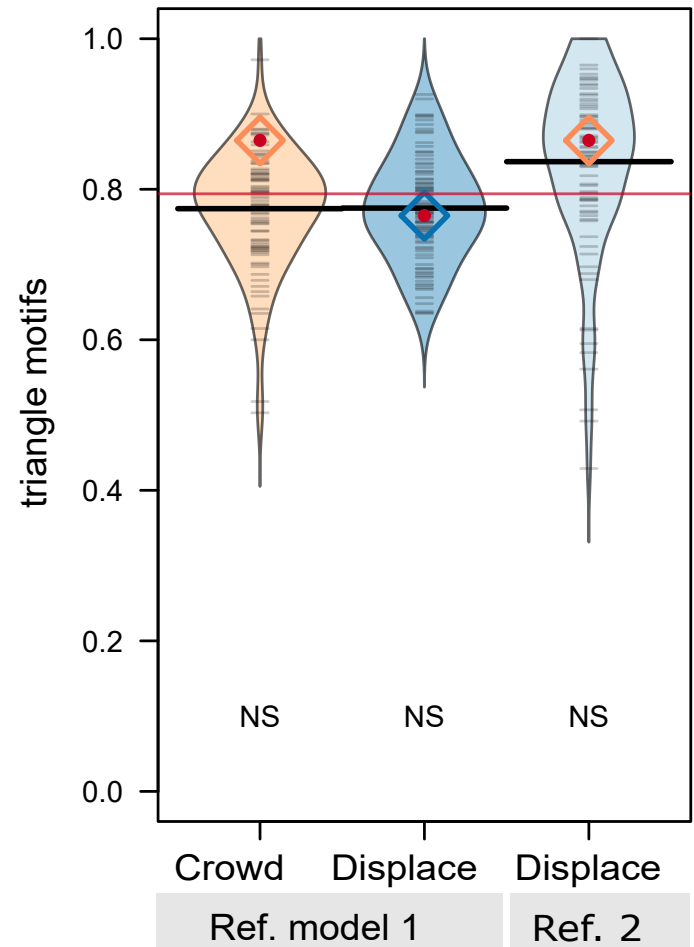

**a) Linearity**

**b) Steepness**

**c) Triangle transitivity**

- Observed value pooled
- Observed value crowd
- Observed value displace
- Ref. model 1 crowd
- Ref. model 1 displace
- Ref. model 2 displace

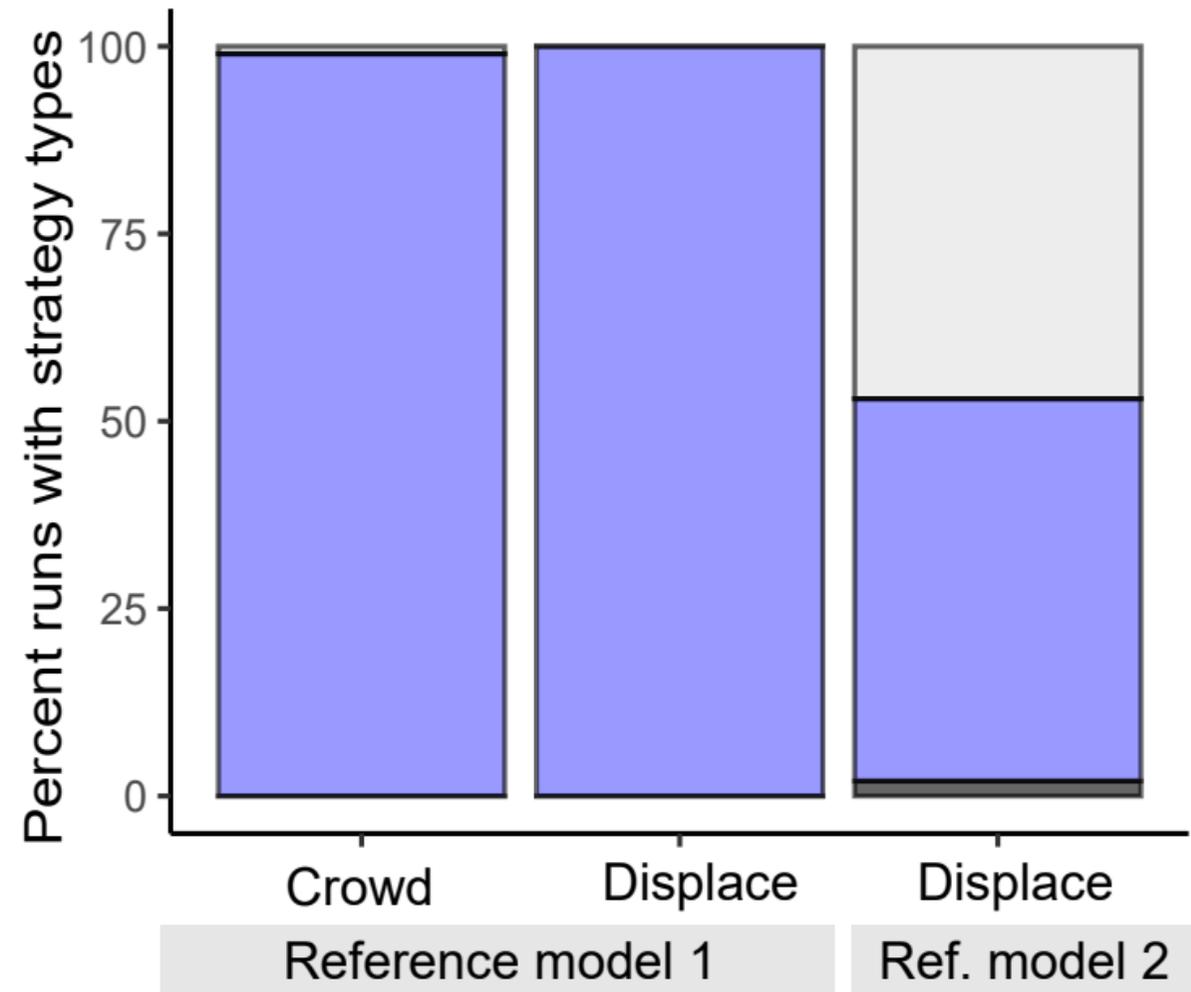

**Supplemental material**

**A framework to evaluate whether to pool or separate behaviors in a multilayer network**

Running head: Decision framework for pooling or splitting behaviors

Annemarie van der Marel1, Sanjay Prasher1, Chelsea Carminito1, Claire O'Connell1, Alexa Phillips1, Bryan M. Kluever2, Elizabeth A. Hobson1

1 Department of Biological Sciences, University of Cincinnati, Ohio USA

2 United States Department of Agriculture, Wildlife Services, National Wildlife Research Center, Florida Field Station, Florida USA

### *Supplemental material 1: reference models*

We developed two reference models to analyze the implications of pooling or splitting two behaviors within a behavioral context. We refer to a reference model when we randomized some features of the data but preserved other parts. We developed reference model 1 to address whether two agonistic behaviors (crowds and displacements) are interchangeable (schematic diagram Figure SM 1). We constructed reference model 2 to address whether the results from reference model 1 are due to unequal rate of observations (crowds n = 160, displacements n = 1055; schematic diagram Figure SM 2).

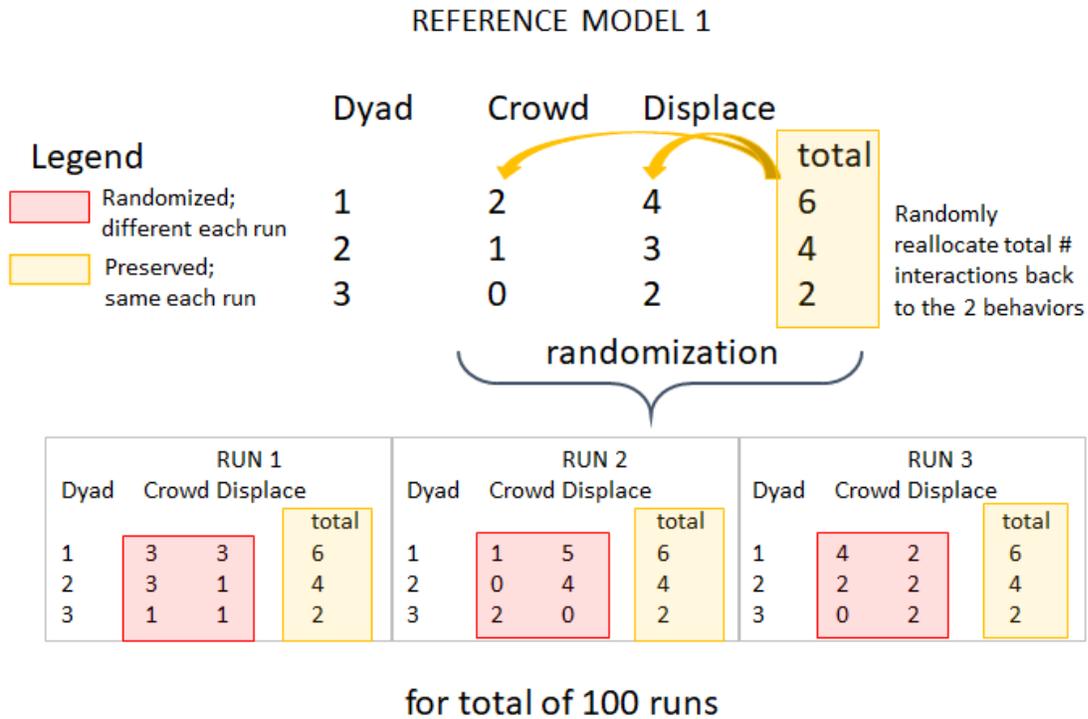

**Figure SM 1. A schematic diagram of the randomization process of reference model 1**

In reference model 1, we randomly reallocated the total number of interactions back to the crowds and displacements. This reference model preserves the total number of individuals in the group, which individuals interact in an agonistic context, and number of total agonistic interactions. The reference model randomizes only the number of interactions that are categorized as displacements versus crowds (n=100 runs).

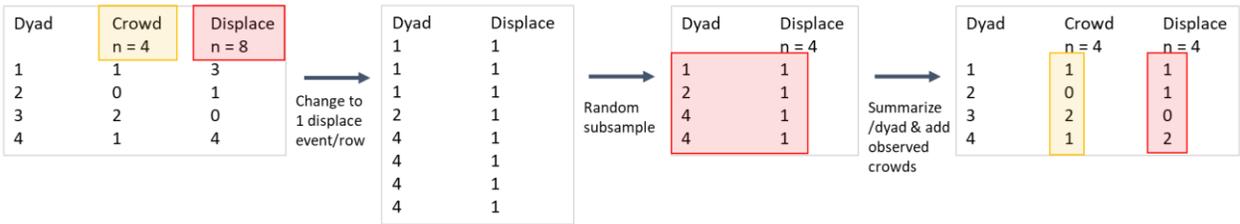

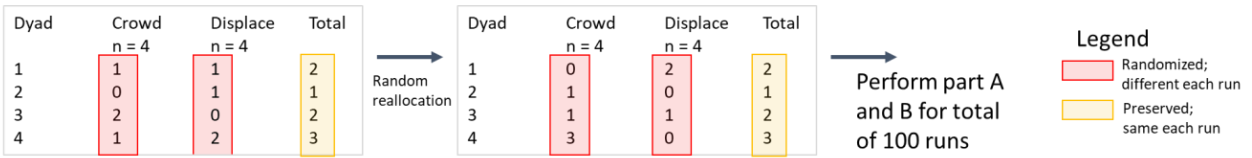

**Figure SM 2. A schematic diagram of the randomization process of reference model 2**

In reference model 2, we subsampled the displacement data to equal the number of interactions of crowds (Part A).This reference model preserves the total number of individuals in the group, which individuals interacted in each agonistic context (crowd versus displacements), and the number of crowds observed for each dyad. The reference model randomizes only which of the total observed displacements were subsampled. We then summarized the total of the observed crowds and the subsampled displacements and randomly re-allocated the total back to the two interaction types for 100 runs (Part B).